\documentclass{PoS}
\usepackage{graphicx,color}
\newcommand{\bda}{\begin{\displaymath}\begin{array}{rl}}
\newcommand{\eda}{\end{array}\end{displaymath}}
\newcommand{\be}{\begin{equation}}
\newcommand{\ee}{\end{equation}}
\newcommand{\bdm}{\begin{displaymath}}
\newcommand{\edm}{\end{displaymath}}
\newcommand{\bea}{\begin{eqnarray}}
\newcommand{\eea}{\end{eqnarray}}

\newcommand{\fs}{\; \; .}

\newcommand{\indR}{\mbox{\scriptsize R}}
\newcommand{\indL}{\mbox{\scriptsize L}}
\newcommand{\qbar}{\overline{\rule[0.42em]{0.4em}{0em}}\hspace{-0.45em}q}
\newcommand{\ubar}{\overline{\rule[0.42em]{0.4em}{0em}}\hspace{-0.5em}u}
\newcommand{\dbar}{\,\overline{\rule[0.65em]{0.4em}{0em}}\hspace{-0.6em}d}
\newcommand{\sbar}{\overline{\rule[0.42em]{0.4em}{0em}}\hspace{-0.5em}s}
\newcommand{\Dbar}{\overline{\rule[0.65em]{0.5em}{0em}}\hspace{-0.65em}D}

\newcommand{\lvac}{\langle 0|\,}
\newcommand{\rvac}{\,|0\rangle}

\newcommand{\lambdabar}{\lambda\hspace{-0.57em}\rule[0.5em]{0.4em}{0.05em}}
\def\eqref#1{(\ref{#1})}
\def\text#1{\mbox{\tiny #1}}

\title{Theoretical aspects of Chiral Dynamics}

\ShortTitle{Theoretical aspects of Chiral Dynamics}

\author{\speaker{H.~Leutwyler} \\
       Albert Einstein Center for Fundamental Physics\\ Institut f\"ur theoretische Physik, Universit\"at Bern, Sidlerstr. 5, CH-3012 Bern, Switzerland\\
       E-mail: \email{Leutwyler@itp.unibe.ch}}

\abstract{Many of the quantities of interest at the precision frontier in particle physics require a good understanding of the strong interaction at low energies. The present talk reviews the theoretical framework used in this context. In particular, I draw attention to the fact that applications of effective field theory methods in the low energy domain involve two different aspects: dependence of the quantities of interest on the quark masses and dependence on the momenta. While the lattice approach gives an excellent handle on the low energy constants that govern the  quark mass dependence, the most efficient tool for pinning down the momentum dependence is dispersion theory. At the same time, the dispersive analysis  enlarges the energy range where the effective theory applies. In the meson sector, the interplay of the various sources of information has led to a coherent framework that describes the low energy structure at remarkably high resolution. The understanding of the low energy properties in the baryon sector is less well developed.  There is significant progress in the dispersive analysis of $\pi N$ scattering, for example, but it leads to puzzling conclusions concerning the pattern of SU(3) symmetry breaking in the baryon octet, which yet remain to be understood. Finally, I critically examine recent papers dealing with the Cottingham formula for the electromagnetic contribution to the mass difference between proton and neutron.}

\FullConference{The 8th  International Workshop on Chiral Dynamics \\
		 29 June 2015 - 03 July 2015\\
		 Pisa, Italy}

\begin{document}

\section{Introduction}
At low energies, the lightest particles play the most important role. The lightest strongly interacting particles are the pions.  We know why they are so light: they represent the Nambu-Goldstone bosons of a hidden internal symmetry. For the analysis of the low energy structure of QCD, this symmetry plays an essential role, because it very strongly constrains the properties of the pions.  

Almost immediately after the discovery of the neutron \cite{Chadwick}, Heisenberg pointed out that the approximate equality of the proton and neutron masses can be understood if the strong interaction is assumed to have an internal symmetry: isospin symmetry \cite{Heisenberg}. Indeed, for a long time, it was taken for granted that this symmetry is an exact property of the strong interaction and that the electromagnetic interaction is responsible for the mass difference. This looks plausible, because the electromagnetic interaction roughly has the proper strength, but it leads to a puzzle: despite the fact that the energy stored in the electric field surrounding the proton increases the mass rather than lowering it, the neutron is the heavier one of the two. 

The puzzle was solved only in 1975, when it was realized that QCD can describe the strong interaction correctly only if $m_u$ is very different from $m_d$, i.e.\,if this interaction breaks isospin symmetry \cite{GL 1975}. The crude estimates for the ratios of the three lightest quark masses obtained in that work, $m_u/m_d\simeq 0.67$, $m_s/m_d\simeq 22.5$, have in the meantime been improved considerably. In particular, Weinberg~\cite{Weinberg 1977} pointed out that in the chiral limit, the Dashen theorem \cite{Dashen} provides an independent estimate of the quark mass ratios, as it determines the electromagnetic self-energies of the kaons in terms of those of the pions. Neglecting higher orders in the expansion in powers of $m_u,m_d$, and $m_s$, he obtained the estimate $m_u/m_d\simeq 0.56$, $m_s/m_d\simeq 20.1$. Also, the decay $\eta\rightarrow 3\pi$ turned out to be a very sensitive probe of isospin breaking~\cite{Gasser:1984pr,Kambor:1995yc,Anisovich:1996tx,Bijnens:2007pr,Schneider:2010hs,Colangelo:2011zz,Albaladejo:2015bca}. The quark mass ratios obtained from that source also confirm the picture.  According to the most recent edition of the FLAG review~\cite{FLAG2014}, the current lattice averages are $m_u/m_d=0.46(3)$, $m_s/m_d= 20.0(5)$ (for a review of the current state of the art on the lattice, see the talk by Claude Bernard \cite{Bernard talk}).

\section{Chiral symmetry}

The resolution of the puzzle gives rise to a new one: since $m_u$ is very different from $m_d$ -- how come that, nevertheless, isospin is a nearly perfect symmetry ? The explanation also relies on symmetry, more precisely on the hidden symmetry of the strong interaction discovered by Nambu \cite{Nambu}, even before the advent of QCD. 

When studying the properties of the weak axial current, Nambu concluded that (1) the strong interaction must have an approximate chiral symmetry and (2) a phenomenon known to occur in solid state physics (magnets, superconductors) must also take place in particle physics. The phenomenon originates in the fact that the symmetry of the Lagrangian does not guarantee that the state of lowest energy is symmetric. If the Lagrangian is symmetric and the ground state is asymmetric, then the spectrum of the theory necessarily contains massless bosons. Nowadays, these particles are called Nambu-Goldstone bosons and symmetries of the Lagrangian that are not shared by the ground state are referred to as hidden or spontaneoulsy broken. If the Lagrangian is only approximately symmetric, so that the currents related to the generators of the symmetry group are not strictly conserved, the spectrum does not contain massless bosons, but particles with a small mass: the energy gap between the ground state and the first excited state does not vanish, but must be small. 
Nambu realized that the pions are the approximately massless particles generated by the spontaneous breakdown of an approximate chiral symmetry. 

In QCD, the presence of an approximate chiral symmetry is not mysterious at all: it so happens that $m_u$ and $m_d$ are very small. In the {\it chiral limit}, where the two masses are set equal to zero, QCD becomes invariant under independent flavour rotations of the right- and left-handed $u,d$-fields. The corresponding symmetry group is SU(2)$_{\indR}\times$SU(2)$_{\indL}$.  Isospin symmetry, SU(2)$_{\indR + \indL}$, is a subgroup thereof and hence becomes exact in the chiral limit.

\section{Mass of the pion}

For $m_u=m_d=0$, QCD acquires an exact SU(2)$_{\indR}\times$SU(2)$_{\indL}$ symmetry. In that limit, the pions represent the Nambu-Goldstone bosons of an exact hidden symmetry and hence are strictly massless.  Gell-Mann, Oakes and Renner \cite{GMOR} pointed out that, for small values of $m_u,m_d$, the square of the mass of the charged pion is proportional to $m_u+m_d$:
\be\label{eq:GMOR} M_{\pi^+}^2=(m_u+m_d)\times|\lvac \ubar u\rvac|\times 
\frac{1}{F_\pi^{2}}\,,\ee  
where $F_\pi=92.28(9)$ MeV \cite{Rosner Stone} is the pion decay constant. The relation states that the pion mass is determined by the geometric mean of the quantity $m_u+m_d$, which measures the breaking of chiral symmetry in the Lagrangian of the theory and the quark condensate $|\lvac \ubar u\rvac|$, which measures the asymmetry of the ground state (since the transformation law of the operator $\ubar u$ under independent isospin rotations of the left- and right-handed components of the quark field does not contain a singlet, it can have a nonzero vacuum expectation value only if the vacuum is not invariant). 

If the electroweak interactions are switched off, the pion mass is determined by the parameters that characterize QCD:
\be \label{eq:MpiQCD}M_{\pi^+}^2=M_{\pi^+}^2(\Lambda_{\mbox{\tiny QCD}},m_u,m_d,m_s,m_c,m_b,m_t)\,.\ee
The Gell-Mann-Oakes-Renner formula \eqref{eq:GMOR} states that the expansion of this function in powers of $m_u$ and $m_d$ (all other parameters being kept fixed at their physical values) starts with a term linear in $m_u$ and $m_d$. 
Remarkably, only the sum of the two quark masses counts and the leading terms in the expansion of $M_{\pi^+}^2$ and $M_{\pi^0}^2$ are exactly the same -- the difference between the squares of the charged and neutral pion masses is of order $(m_d-m_u)^2$ and thus only shows up if the expansion is taken beyond first order. 

These properties reflect the fact that isospin symmetry is not hidden: in the chiral limit, the ground state is invariant under isospin rotations. In fact, the entire leading term in the Lagrangian of chiral perturbation theory ($\chi$PT) fails to take notice of $m_d-m_u$: the Nambu-Goldstone bosons are protected from isospin breaking (i.e.~from the part of the Lagrangian that is not invariant under  the subgroup which is not spontaneously broken). This, finally, explains why isospin is a nearly perfect symmetry of nature, despite the fact that $m_u$ is very different from $m_d$.

The work done on the lattice yields a beautiful confirmation of the Gell-Mann-Oakes-Renner formula and shows that the linear term in the expansion of $M_{\pi^+}^2$ dominates out to values of $m_u$ and $m_d$ that are about ten times larger than in nature \cite{Duerr}. The lattice data also allow a determination of the condensate, currently to an accuracy of 10 or 20 \% \cite{FLAG2014}. 

\section{Higher orders of the chiral expansion}
At the next order, the expansion of the pion mass in powers of the quark masses contains a chiral logarithm. If $m_u$ is set equal to $m_d$, the representation obtained by evaluating $\chi$PT to one loop reads 
\be\label{eq:Mpiloop}M_{\pi^+}^2= M^2\left\{1- \frac{M^2}{2(4\pi F)^2}\bar{\ell}_3+O(M^4)\right\}\,,\hspace{0.5cm}\bar{\ell}_3=\ln\frac{\Lambda_3^2}{M^2}\,,\ee
where $M^2$ stands for the term linear in the quark masses,\footnote{$F$ and $|\lvac \ubar u\rvac|$ represent the pion decay constant and the condensate in the chiral limit.} 
\be\label{eq:M} M^2\equiv (m_u+m_d)|\lvac \ubar u\rvac|\frac{1}{F^2}\,.\ee
Chiral symmetry does not determine the scale $\Lambda_3$. This  scale fixes the value of the corresponding low energy constant (LEC) $\ell_3$, which depends on the running scale $\mu$ at which the loop graphs of $\chi$PT are renormalized: $\ell_3(\mu)=-\ln (\Lambda_3^2/\mu^2)/64\pi^2$. The lattice data show clear evidence for the presence of a term that logarithmically depends on the quark masses. The numerical value quoted in \cite{FLAG2014} is $\bar{\ell}_3$ = 3.05(99) for $M$ = 135 MeV, indicating that the scale of the logarithm is of order $\Lambda_3\simeq$ 600 MeV. In view of the factor $(M/4\pi F)^2$ in front of the logarithm, the correction is tiny: at the physical value of the quark masses, it amounts to about 2.4 \%. This illustrates the fact that the quark masses $m_u$ and $m_d$ are very small -- SU(2)$_{\indR}\times$SU(2)$_{\indL}$ is a nearly perfect hidden symmetry of QCD: the relevant symmetry breaking parameter, $m_u+m_d$, is only about three times larger than the difference $m_d-m_u$, which measures the strength of isospin breaking by the strong interaction.

\section{Interaction among the pions}

If the electroweak interaction is switched off and $m_u$ is set equal to $m_d$, isospin symmetry becomes exact. It implies that the scattering of any of the 6 initial states $\pi^+\pi^+,\pi^+\pi^0, \pi^+\pi^-,\,\ldots$ into any of these final states is described by a single function $A(s,t)$. As pointed out by Weinberg, almost 50 years ago \cite{Weinberg scattering lengths}, current algebra implies that the expansion of this amplitude in powers of the momenta and of $m_u=m_d=\hat{m}$ starts with 
\be\label{eq:ALO} A(s,t)=\frac{1}{F_\pi^2}(s-M_\pi^2)+\ldots \ee
In other words, chiral symmetry implies a parameter free prediction for the strength of the interaction, valid to leading order in the expansion in powers of momenta and quark masses. The formula shows that, in the chiral limit, where the pions are massless particles, the scattering amplitude vanishes at zero momentum. For the S-wave scattering lengths,\footnote{These formulae represent the scattering lengths in units of $M_\pi$. The conventional scattering lengths are obtained by multiplying the pure numbers in equation \eqref{eq:a0a2LO} with the reduced pion Compton wavelength, $\lambdabar_\pi=\hbar/M_\pi c$. Since the factor $\lambdabar_\pi$ diverges in the chiral limit, it distorts the chiral power counting and is omitted in these formulae.}  the formula \eqref{eq:ALO} implies \cite{Weinberg scattering lengths}
\be\label{eq:a0a2LO}
a_0 =  \frac{7M_\pi^2}{32\pi F_\pi^2} = 0.16\,,\hspace{1cm}a_2 = -\frac{M_\pi^2}{16\pi F_\pi^2} = -0.045\,.\ee
The expressions represent the leading terms in the expansion of $a_0,a_2$ in powers of $\hat{m}$. They show that, at threshold, the interaction is attractive in the channel with $I=0$, repulsive for $I=2$ and disappears in the chiral limit.  

The chiral perturbation series of the $\pi\pi$ scattering amplitude has been worked out to NNLO \cite{pipiNNLO1,pipiNNLO2}. While the exotic scattering length $a_2$ practically stays put, the corrections in $a_0$ are surprisingly large: the NLO corrections increase $a_0$ by 26\% and those of NNLO generate a further enhancement of about 8\%. That seems to contradict the statement that SU(2)$_{\indR}\times$SU(2)$_{\indL}$ is a nearly perfect symmetry of the strong interaction -- how come that, in the case of the scattering lengths, the convergence of the expansion in powers of the symmetry breaking parameter $\hat{m}$ is so slow ?

The reason is that the interaction among the Nambu-Goldstone bosons is weak only at low energies -- it very rapidly grows with the energy. The partial wave amplitude of the isoscalar S-wave, $t_0(s)$, clearly exhibits this behaviour. The leading order contribution is readily obtained from \eqref{eq:ALO} and reads $t_0(s)=(2s-M_\pi^2)/32 \pi F_\pi^2$. This expression has an Adler zero at $s=\frac{1}{2}M_\pi^2$, but linearly rises, reaching the value quoted in \eqref{eq:a0a2LO} at threshold ($s=4M_\pi^2$), where $t_0(s)$ represents the scattering length. Unitarity generates a branch point there. The singularity produces curvature and strongly bends the amplitude upwards, amplifying the value of $a_0$. This is reflected in the chiral perturbation series of $a_0$, which contains a juicy chiral logarithm at NLO: 
\be a_0=\frac{7M_\pi^2}{32\pi F_\pi^2}\left\{1+\frac{9M_\pi^2}{2(4\pi F_\pi)^2}\ln\frac{\Lambda_0^2}{M_\pi^2}+O(M^4)\right\}\,.\ee
The comparison with \eqref{eq:Mpiloop} shows that the coefficient of the chiral logarithm in $a_0$ is nine times larger than the one occurring in $M_\pi^2$.

 The essential point of the above discussion is that $\chi$PT not only involves an expansion in powers of the quark masses, but also one in powers of the momenta. While the expansion of $M_\pi^2$ exclusively concerns the dependence on the quark masses, the scattering amplitude also depends on the momenta. It is clear that the accuracy to which the momentum dependence is accounted for by the first few terms of the chiral perturbation series depends on the magnitude of the momenta considered, which unlike the quark masses represent free variables that are not determined by QCD. The above discussion shows that, at threshold, the leading term of the chiral series does represent the dominating contribution, but the higher orders generate corrections that are much larger than those seen in the expansion of $M_\pi^2$. 
 
\section{Dispersion theory}\label{sec:Dispersion theory}
The slow convergence of the chiral expansion encountered in the case of the scattering length $a_0$ does not arise from the expansion in powers of $\hat{m}$, but from the one in powers of the momenta. Actually, $\chi$PT is not needed to determine the dependence on the momenta. Dispersion theory is a much more efficient tool for that. 

As shown by Roy \cite{Roy}, analyticity, unitarity and crossing symmetry very strongly constrain the $\pi\pi$ scattering amplitude. In this framework, the S-wave scattering lengths $a_0,a_2$ enter as subtraction constants. In \cite{ACGL}, the Roy equations are solved numerically. The dispersion integrals are split into a low energy region $4M_\pi^2<s<s_0$ and a remainder, $s_0<s<\infty$. The {\it matching point} $s_0$ is taken at $\sqrt{s_0}=800$ MeV. Below that point, the elasticities of the partial waves are treated as known, while the input of the calculation in the high energy region consists of the imaginary parts of the scattering amplitude, which are taken from experiment. As shown in \cite{ACGL,CGL},  the scattering lengths $a_0,a_2$,  the elasticities below and the imaginary parts above the matching point unambiguously determine the scattering amplitude $A(s,t)$ throughout the low energy region -- within the uncertainties generated by the noise in the input.  

\begin{figure}[h]\hspace{15mm} {\footnotesize rapid convergence}\hspace{1cm}{\footnotesize slow convergence\hspace{1cm} of the chiral series}

\hspace{3.2cm}{ $\Downarrow$} \hspace{1.95cm}{ $\Downarrow$} 

\vspace{0.2em}\hspace{2.55cm}\includegraphics[width=6cm]{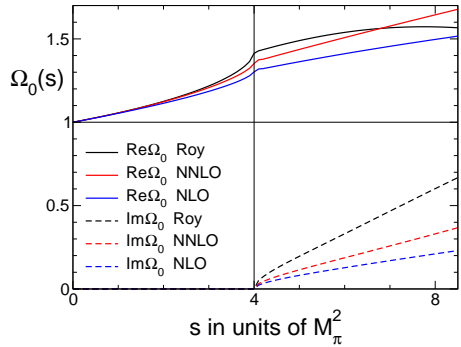}

\caption{\label{fig:Omnes}Comparison of $\chi$PT with dispersion theory: energy dependence of the Omn\`{e}s factor belonging to the isoscalar S-wave.}
\end{figure}

In figure \ref{fig:Omnes}, the output of the dispersive calculation is compared with the chiral representation for the case of the Omn\`{e}s factor $\Omega_0(s)$, which is defined by 
\be\Omega_0(s)=\exp \frac{s}{\pi}\hspace{-0.2em}\int_{4M_\pi^2}^\infty\hspace{-0.2em}\frac{ds'}{s'}\frac{\delta_0(s')}{s'-s-i\epsilon}\,,\ee
 where $\delta_0(s)$ is the phase-shift of the isoscalar S-wave \cite{Omnes}. This function describes the momentum dependence generated by the final state interaction, in the approximation where inelastic transitions are neglected. It plays a central role in the dispersive analysis of form factors and scattering amplitudes.  
The curves in the upper half of the figure show the real part of the Omn\`es factor, those in the lower half represent the imaginary part.\footnote{I thank Peter Stoffer for this plot.} 

The chiral perturbation series starts with $\Omega_0(s)=1+O(q^2)$. The lowest line in the upper half of the figure shows the behaviour of the chiral representation at NLO, the next higher one includes NNLO corrections and the top one indicates the behaviour obtained by solving the Roy equations. On the interval shown in the figure, these equations determine $\Omega_0(s)$ to good precision (for details see \cite{CGL}). 
All three curves show the rapid, approximately linear rise at small values of $s$, as well as the curvature generated by the cusp at $s=4M_\pi^2$. The corrections from the various terms of the chiral series visibly grow with the energy. At $s=4M_\pi^2$, their relative size is comparable with the one seen in the chiral expansion of the scattering lengths, even a little larger. 

In the vicinity of $s=0$, the three curves can barely be distinguished: there, the expansion in powers of $s$ is totally dominated by the constant and linear terms. Since the contributions from the higher powers of momentum are tiny, the chiral series rapidly converges there, like in the case of $M_\pi^2$. This property is made use of in \cite{CGL}, where it is shown that a remarkably accurate prediction for the scattering lengths is obtained by matching the dispersive and chiral representations at $s=0$ (in this language, fixing the subtraction constants $a_0,a_2$ directly with the result obtained within $\chi$PT corresponds to matching the two representations at $s=4M_\pi^2$). Indeed, the speed of the convergence achieved with this method is amazing: the predictions at LO, NLO, NNLO are 0.197, 0.2195, 0.220 for $a_0$ and -0.0402, -0.0446, -0.0444 for $a_2$. 

\section{Comparison with experiment and lattice results}
\label{sec:Comparison}
The example shows that, in contrast to the straightforward expansion in powers of the momenta, which in the case of the scattering amplitude rapidly converges only in the vicinity of the Adler zero, dispersion theory provides a decent description of the momentum dependence, even in the physical region, above threshold. The precision achieved by combining the low energy theorems of chiral perturbation theory with dispersive methods triggered new low energy experiments concerning kaon decays $K\rightarrow e\nu\pi\pi$ \cite{Brookhaven,CERNKell4}, $K\rightarrow 3\pi$ \cite{CERNK3pi} and $\pi^+\pi^-$ atoms \cite{Adeva:2011tc}. First observations of atoms formed with charged kaons and pions -- a fascinating laboratory for the experimental investigation of QCD at low energies -- have also been reported \cite{Adeva:2014xtx}. 

For the properties of hadronic atoms, QED evidently plays a central role \cite{Gasser:2007zt} and for the phenomena observed in the decay 
$K^\pm\rightarrow \pi^\pm\pi^0\pi^0$, in the immediate vicinity of the threshold, the mass difference between the charged and neutral pions, which is predominantly of electromagnetic origin, is also essential \cite{Budini Fonda,Cabibbo:2004gq,Cabibbo:2005ez,Colangelo:2006va}. Even in processes where isospin breaking only generates corrections, these play a significant role at the accuracy reached in some of the experiments and must be accounted for when drawing conclusions from what is observed \cite{Colangelo:2008sm}. Moreover, in order to establish firm contact between experiment and the Standard Model -- sine qua non if evidence for physics beyond this framework is to be found at low energies --  a dispersive analysis of the relevant processes is required. Quite a few talks given at this conference were devoted to these topics -- I cannot review this here, but refer to the corresponding contributions in the present proceedings.  

Today, the lattice approach provides the most precise source of information about the $\pi\pi$ S-wave scattering lengths. The exotic one, $a_2$, can be determined directly from the volume dependence of the energy levels on the size of the box used to formulate QCD on a lattice. Alternatively, and this also works for $a_0$, the dominating low energy constants of $\chi$PT can be determined on the lattice. In the isospin limit and to one loop, the chiral representation of the scattering amplitude involves four LECs: $\ell_1,\ell_2,\ell_3,\ell_4$. While $\ell_1,\ell_2$ concern the momentum dependence, $\ell_3,\ell_4$ determine the dependence on the quark mass $\hat{m}$. As discussed above, dispersion theory  provides accurate information about the momentum dependence. Indeed, if $\ell_3,\ell_4$ are known, the Roy equations can be used to determine $\ell_1, \ell_2$  within narrow limits \cite{CGL}. 

While the constant $\ell_3$ can be extracted from the quark mass dependence of $M_\pi$, the constant $\ell_4$ concerns the dependence of the pion decay constant on the quark mass. The one loop formula analogous to \eqref{eq:Mpiloop} reads
\be\label{eq:Fpiloop}F_\pi = F\left\{1+ \frac{M^2}{(4\pi F)^2}\bar{\ell}_4+O(M^4)\right\}\,,\hspace{0.5cm}\bar{\ell}_4=\ln\frac{\Lambda_4^2}{M^2}\fs\ee 
The constant $\ell_4$ can thus be determined by studying the dependence of the pion decay constant on the mass of the two lightest quarks.  

\begin{figure}[h]
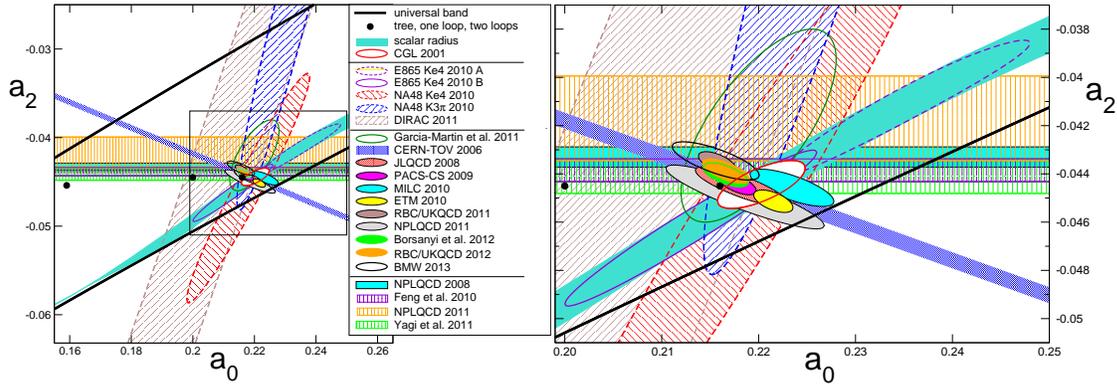

 \includegraphics[width=7.25cm]{a0a2_201510_square.eps}\hspace{-0.2cm}
\raisebox{-0.02cm}{
\includegraphics[width=7.5cm]{a0a2_201510_nolegend.eps}}
\caption{\label{fig:a0a2}Comparison of the predictions for the $\pi\pi$ scattering lengths with experimental and lattice determinations.}
\end{figure}
Figure \ref{fig:a0a2} compares the theoretical prediction for the scattering lengths (white ellipse) with (1) experimental determinations\footnote{While all other data are corrected for isospin breaking, the published result of E865, $a_0=0.235(13)$ \cite{Brookhaven}, is not. According to Table 6 of \cite{CERNKell4}, the  isospin breaking corrections \cite{Colangelo:2008sm} lower this to $a_0=0.213(13)$. The dashed ellipse (E865 A) is obtained by combining the published result with the constraint derived from the scalar radius \cite{CGL}, while the full one (E865 B) accounts for the corrections. The difference between the two shows that, at the precision reached for the scattering lengths, isospin breaking generates a very pronounced effect (see the talk by Marc Knecht \cite{Knecht talk}). } (E865 \cite{Brookhaven}, NA48 Ke4 \cite{CERNKell4}, NA48 K3$\pi$ \cite{CERNK3pi}, DIRAC \cite{Adeva:2011tc}), (2) results for $a_2$ extracted from the volume dependence of the energy levels on the lattice and (3) values for $a_0,a_2$ obtained by combining lattice results for $\ell_3,\ell_4$ with the Roy equation analysis. The right panel focuses on the square indicated in the left panel.  

While at NLO, the quark mass dependence is controlled by $\ell_3,\ell_4$, the analogous contributions occurring at NNLO involve four LECs: $r_1,r_2,r_3,r_4$.  The ellipses obtained with (3), as well as the white one that represents the theoretical prediction, include a crude estimate of these constants, which indicates that their effects are too small to be visible at the accuracy reached. Apart from the fact that some of the lattice collaborations appear to underestimate the systematic errors, the picture is perfectly coherent. The fact that the experimental determination of the scattering lengths agrees with the low energy theorems of Weinberg subjects our understanding of the low energy structure of the Standard Model to a very strong test.  

At higher orders, the formulae relating the masses and decay constants to the parameters occurring in the chiral Lagrangian become complicated. It is of interest to isolate the dominating contributions, approximating the numerical representations of the relevant loop integrals with algebraic expressions. Considerable efforts have been undertaken in this direction \cite{Kaiser:2006uv,Kaiser:2007kf,Ecker:2010nc,Ecker:2013pba,Anant}. 
The lattice is the ideal tool to determine the higher order contributions, in \eqref{eq:Mpiloop} as well as in \eqref{eq:Fpiloop}. As a plea addressed to the lattice community: please do not be content with reaching the physical values of the quark masses, but make the knowledge acquired about the way the investigated quantities depend on the quark masses accessible, i.e.~determine the corresponding LECs. These represent well-defined properties of QCD and play an important role in the low energy analysis. While it is not possible to vary the quark masses experimentally, this does not pose a problem for the lattice approach -- quite to the contrary, accurate values are more easy to obtain if $m_u,m_d$ are taken larger than in nature. 

It makes an essential difference here whether one wishes to determine the LECs of the effective Lagrangian based on SU(2)$_{\indR}\times$ SU(2)$_{\indL}$ or aims at those of  the SU(3)$_{\indR}\times$SU(3)$_{\indL}$-Lagrangian. In the latter case, the effective theory treats not only the pions, but also the kaons and the $\eta$ as approximately massless.  If $m_s$ is held fixed at the physical value, while $m_u,m_d$ are taken significantly heavier than in nature,  then the kaon and $\eta$ masses become too large for the first few terms of their chiral expansion to represent a decent approximation. While pion masses of order 300 or 400 MeV are within the range where SU(2)$_{\indR}\times$SU(2)$_{\indL}$ does provide a coherent framework, a meaningful determination of the LECs of SU(3)$_{\indR}\times$SU(3)$_{\indL}$ requires data in a range where the entire pseudoscalar octet is light  -- for the chiral representation to be accurate and the nonleading terms to be visible, the meson masses should be neither too large nor too small.
 
Many processes have been analyzed within the effective theory, quite a few even to NNLO. In particular, Hans Bijnens and his group provide explicit representations for many quantities of physical interest \cite{Bijnens}. Also, the leading chiral logarithms have been worked out for many observables in the meson sector and first results in the nucleon sector are also available (see the talk by Hans Bijnens \cite{Bijnens talk}). For a review of the status of $\chi$PT in the meson sector, in particular also for a discussion of the current knowledge of the low energy constants, I refer to the talk of Gerhard Ecker \cite{Ecker talk}.

\section{Developments in dispersion theory}
Early studies of the low energy structure of the Standard Model clearly revealed the presence of resonances such as $\rho,\omega,\phi,\ldots$ In the region below the $\rho$, however, the situation was far from clear. There were indications for the occurrence of a resonance in the channel with $I=J=0$, now referred to as $f_0(500)$, but the analysis invariably involved extrapolations and led to quite a spread in the outcome for mass and width. Even the very existence of this resonance was disputed -- for a thorough review of the history, I refer to \cite{Pelaez:2015qba}. 

The Roy equations put the dispersive analysis of the low energy structure on solid mathematical grounds. In particular, in the egg-shaped region ${\cal E}$ of the complex $s$-plane where these equations are valid, the interaction among the pions can be calculated in a controlled manner. As demonstrated in \cite{CCL}, the partial wave amplitude with $I=J=0$ contains a pair of conjugate zeros in ${\cal E}$ and, on account of unitarity, a pair of poles at the same place on the second sheet. The position of the poles on the second sheet determines mass and width of the resonance, which can thus be calculated in a straightforward manner (within the uncertainties attached to the input, but these have a remarkably small effect on mass and width). The work done since then fully confirmed the result \cite{Pelaez:2015qba}. Moreover, the analysis was extended to the $K\pi$ channel, with the result that the existence of the $K_0^\star(800)$ is now also established beyond doubt -- mass and width have been calculated to an accuracy comparable to the one reached for the $f_0(500)$ \cite{DescotesGenon:2006uk}.

While $\chi$PT provides a useful representation of the scattering amplitude only in the unphysical region below threshold, the range of validity of the Roy equations extends beyond a centre-of-mass-energy of 1 GeV. The solution of these equations yields an explicit representation for all of the partial waves. In particular, it accurately describes the most prominent low energy phenomenon, the $\rho$-resonance, not only in the vicinity of the resonance peak (where the Breit-Wigner approximation works quite well because the pole sits close to the real axis), but also on the wings of the resonance.  

Let me draw attention to a puzzling discrepancy between recent results obtained on the lattice \cite{Bai:2015nea} and the dispersive analysis in \cite{CGL}. It concerns the $\pi\pi$ phase shifts relevant for the decay $K\rightarrow \pi\pi$. In the Standard Model, the phase of $\epsilon'/\epsilon$ is given by the difference $\delta_0-\delta_2$ between the S-wave phase shifts  at $\sqrt{s}=M_{K^0}$ (the index refers to isospin, which can take the values $I=0,2$).  Using L\"uscher's quantization condition, the RBC/UKQCD collaboration arrives at  $\delta_0 = 23.8(4.9)(1.2)^\circ$ and $\delta_2 =-11.6(2.5)(1.2)^\circ$. The corresponding central value of the difference, $\delta_0-\delta_2=35.4^\circ$, is far outside the range permitted by the uncertainties in the prediction obtained from the Roy equations, $\delta_0-\delta_2=47.7(1.5)^\circ$ \cite{CGL}. A confirmation of the lattice result would lead to an Aha!-experience of first rank: I can see no way to accommodate a phase shift difference as low as this in the Roy analysis. 

Very significant progress has been made in the dispersive analysis of the form factors relevant for $K_{\ell4}$ decay \cite{Colangelo:2015kha}. At the precision required to look for effects beyond the Standard Model, isospin breaking must be accounted for \cite{Stoffer:2013sfa} (see the talk by Peter Stoffer \cite{Stoffer talk 1}). Furthermore, the contribution to the muon magnetic moment generated by hadronic light-by-light scattering has now been analyzed within dispersion theory \cite{Colangelo:2015ama}. The new analysis provides the basis for a systematic evaluation of this contribution, which currently limits the precision of the theoretical prediction within the Standard Model. For a detailed discussion of this work, I refer to the talks of Gilberto Colangelo \cite{Colangelo talk} and Peter Stoffer \cite{Stoffer talk 2}.  

A qualitatively different development involves advanced dispersive techniques. I cannot review this impressive body of work here, but mention a few illustrative examples: (i) the method leads to bounds for the form factor that describes the low energy contribution from vacuum polarization to the prediction for the muon magnetic moment \cite{Ananthanarayan:2013zua}, (ii) strong constraints on the form factors relevant for $K_{\ell3}$ decay can be established in this way \cite{Abbas:2010ns} and (iii) the approach also provides  stringent consistency tests concerning the $\omega\pi$ transition form factor, which should help resolving the discrepancies between theoretical calculations and some of the data on the process $\omega\rightarrow\pi^0\gamma^\star$ \cite{Ananthanarayan:2014pta,Caprini:2015wja}. For details I refer to the quoted references and, concerning the last topic, to the talk of Balasubramanian Ananthanarayan \cite{Anant talk}. 

\section{$\sigma$-term}
There is very significant progress in the dispersive analysis of $\pi N$ scattering, based on the Roy-Steiner equations \cite{Hite:1973pm,Hoferichter:2009ez,Ditsche:2012fv,Hoferichter:2012wf,Hoferichter:2015dsa,Hoferichter:2015tha, Hoferichter:2015hva}. For a detailed discussion of this work, I refer to the talks of Bastian Kubis and Jacobo Ruiz de Elvira \cite{Kubis  Ruiz talk}. In the following, I limit myself to a few remarks concerning one of the results obtained in this framework -- the value of the $\sigma$-term -- and, moreover, disregard isospin breaking effects, that is set $e=0$, $m_u=m_d=\hat{m}$.

The $\sigma$-term represents the nucleon expectation value of the part of the QCD Hamiltonian that explicitly breaks chiral SU(2)$_{\indR}\times$SU(2)$_{\indL}$ symmetry:
\be \label{eq:sigma}\sigma_N=\frac{\hat{m}}{2M_N}\langle N(p)|\,  \ubar u +   \dbar d |N(p)\rangle\,.\ee
The state $|N(p)\rangle$ describes a nucleon of four-momentum $p$ (the spin direction is not indicated explicitly -- the expectation value is independent thereof). According to the Feynman-Hellman theorem, the matrix element represents the derivative of the nucleon mass with respect to the quark mass, $\sigma_N=\hat{m}\,\partial M_N/\partial \hat{m}$.  
 
The $\sigma$-term has a long history, as it concerns one of the earliest low energy theorems established on the basis of current algebra \cite{Weinberg scattering lengths,Cheng Dashen,Brown Pardee Peccei}. The theorem involves the value of the isospin even $\pi N$ scattering amplitude $D^+\hspace{-0.15em}(\nu,t)$ at the Cheng-Dashen point ($\nu=0$, $t=2M_\pi^2$):  
\be \label{eq:Sigma} \Sigma_N=F_\pi^2\;\Dbar^+\hspace{-0.2em}(0,2M_\pi^2)\ee
(the bar indicates that the contribution from the Born term is removed). The theorem states that, to first order in the expansion in powers of $m_u$ and $m_d$,  there is no difference between $\Sigma_N$ and $\sigma_N$: the scattering amplitude $\Dbar^+(0,2M_\pi^2)$ vanishes in the chiral limit and the first terms in the chiral expansion of  $\Sigma_N$ and $\sigma_N$ are the same.

There is an analogous low energy theorem also for $\pi\pi$-scattering, where -- up to higher order contributions -- the matrix element $\langle \pi(p)| m_u\, \ubar u + m_d\, \dbar d |\pi (p)\rangle$ represents the square of the pion mass: in that case, the entire mass of the particle is due to the breaking of chiral symmetry and the low energy theorem not only relates the scattering amplitude to the $\sigma$-term, but also connects this term with the mass of the particle. In the case of $\pi N$ scattering, the low energy theorem also follows from the fact that the QCD Lagrangian has an approximate SU(2)$_{\indR}\times$SU(2)$_{\indL}$-symmetry, but this symmetry does not predict the value of $\sigma_N$. The difference to the case of the pion also shows up in the size of the corrections: while the chiral expansion of mesonic matrix elements only involves integer powers of the light quark masses and logarithms thereof, the expansion of the baryonic matrix elements goes with powers of the square root of $\hat{m}$. At NLO, the chiral expansion of $\Sigma_N$ as well as the one of $\sigma_N$ picks up contributions from the one loop graphs of $\chi$PT which grow in proportion to $M_\pi^3\sim \hat{m}^{3/2}$ when the quark masses are turned on. 

As pointed out in \cite{Brown Pardee Peccei}, the size of the higher order contributions to the difference $\Sigma_N-\sigma_N$ is reduced if the $\sigma$-term matrix element is evaluated at the same momentum transfer as the scattering amplitude. The matrix element of the quark mass term between nucleons of momentum $p'$ and $p$ is described by the scalar form factor $\sigma(t)$,
\be \langle N(p')| \hat{m}\, (\ubar u +\dbar d) |N(p)\rangle =\sigma(t)\bar{u}(p')u(p)\,,\ee
where $\bar{u}(p')$, $u(p)$ are the corresponding Dirac spinors and $t=(p'-p)^2$. The $\sigma$-term represents the value at the origin: $\sigma_N=\sigma(0)$. 
If the low energy theorem is written in the form
\be\label{eq:LET} \Sigma_N=\sigma(2M_\pi^2)+\Delta_R\,,\ee    
the chiral expansion of the contribution from higher orders, $\Delta_R$, does not start at $O(\hat{m}^{3/2})$, but only at  $\Delta_R=O(\hat{m}^2)$ and is therefore expected to be small. Estimates obtained from resonance exchange in the framework of heavy baryon $\chi$PT lead to $|\Delta_R|\hspace{0.25em}\raisebox{0.15em}{<}\hspace{-0.6em}\raisebox{-0.25em}{$\sim$}\hspace{0.2em} 2\,\mbox{MeV}$ \cite{Bernard:1996nu}. 

The $t$-dependence of the $\sigma$-term, as well as the one of the scattering amplitude $\,\Dbar^+\!(0,t)$, have been investigated in detail \cite{Hoehler,Gasser:1990ce}. The Roy-Steiner analysis referred to above provides a thorough update. It relies on input for the $\pi N$ phase shifts and scattering lengths. For the former, the parameterization of SAID \cite{SAID} is used. The data base that underlies this representation includes recent $\pi N$ cross section measurements and thus goes substantially beyond the data analyzed by the Karlsruhe-Helsinki collaboration used in earlier work on the $\sigma$-term. The scattering lengths are based on the results obtained from the measurements of energy levels and life-times of pionic atoms \cite{Gasser:2007zt}. Note that, since the Born term dominates the amplitude in the low energy region, a reliable determination of the $\pi N $ coupling constant is required. The value used in the Roy-Steiner analysis,  $g_{\pi N}^2/4\pi=13.7(2)$ \cite{Baru:2011bw}, relies on the Goldberger-Miyazawa-Oehme sum rule \cite{Goldberger:1955zza}, more precisely on the evaluations of this sum rule reported in \cite{Ericson:2000md,Abaev:2007nq}.
The value is  consistent with earlier determinations \cite{Pavan:2001wz,Matsinos:2006sw,Sainio:2009zz}, but significantly lower than the result obtained by the Karlsruhe-Helsinki collaboration, $g_{\pi N}^2/4\pi=14.3(2)$ \cite{Hoehler}. For a detailed discussion of the problems encountered in the determination of $g_{\pi N}$, I refer to \cite{Ericson:2000md,Abaev:2007nq}.

The result for the $\sigma$-term can be expressed in terms of the coefficients $d_{00}^+$ and $d_{01}^+$ of the so-called subthreshold expansion:
\be\label{eq:Sigmad} \sigma_N=\Sigma_d+\sigma_N^{\hspace{-0.15em}\mbox{ \tiny rest}}\,,\hspace{3em}\Sigma_d\equiv F_\pi^2(d_{00}^++2M_\pi^2d_{01}^+)\,.\ee
The estimate for the remainder, $\sigma_N^{\hspace{-0.15em}\mbox{ \tiny rest}}=1.2(3.0)\,\mbox{MeV}$ \cite{Hoferichter:2015dsa,Hoferichter:2015tha,Hoferichter:2015hva}, indicates that the subthreshold coefficients represent the crucial quantities in determinations of the $\sigma$-term. While the Karlsruhe-Helsinki analysis led to $d_{00}^+=-1.46(10)\,M_{\pi}^{-1}$, $d_{01}^+=1.14(2)\,M_{\pi}^{-3}$, the outcome of the Roy-Steiner analysis reads $d_{00}^+=-1.36(3)\,M_{\pi}^{-1}$, $d_{01}^+=1.16(2)\,M_{\pi}^{-3}$. Although these numbers are consistent within errors, taken together with the change in the estimate for $\sigma_N^{\hspace{-0.15em}\mbox{ \tiny rest}}$, they imply
\be\label{eq:sigmaRS}\sigma_N=59.1(3.5)\,\mbox{MeV}\,\cite{Hoferichter:2015dsa}\,,\ee
significantly higher than the old estimate $\sigma_N\simeq 45\,\mbox{MeV}$ \cite{Gasser:1990ce}. About 60\%  of the difference come from the change in the subthreshold coefficients, the remainder is due to a difference in the estimate for $\sigma_N^{\hspace{-0.15em}\mbox{ \tiny rest}}$, in particular to isospin breaking effects, which were not accounted for in \cite{Gasser:1990ce}.  

\section{Theoretical estimate for $\sigma_N$}

 From a theoretical viewpoint, the numerical value of the $\sigma$-term obtained from the analysis of data on $\pi N$ scattering is puzzling, because it is in conflict with two assumptions that are part of the generally accepted qualitative understanding of the strong interaction:
 
\vspace{0.4em}
\noindent\hspace{1em}\parbox{14.8cm}{\hspace{-0.8em}$\bullet\,$ SU(3) is a decent approximate symmetry, also for the matrix elements of the operator $\bar{q} \lambda^{\hspace{-0.15em}a} q$ in the baryon octet.

\hspace{-0.8em}$\bullet\,$ The rule of Okubo, Zweig and Iizuka \cite{OZI} is approximately valid. } 
 
 \vspace{0.4em}\noindent
A value around 60 MeV implies that at least one of these two assumptions fails. 

To explain why this is so, I first note that, in the isospin limit, the part of the QCD Lagrangian that breaks SU(3) flavour-symmetry is proportional to the octet operator $\qbar\lambda^{\hspace{-0.2em}8} q\propto\ubar u+\dbar d-2\sbar s$. To first order in the symmetry breaking parameter $m_s-\hat{m}$, the shifts of the baryon masses are given by the expectation values of this operator. For an operator that transforms according to the octet representation and is sandwiched between two octets of physical states, SU(3) symmetry allows only two independent couplings. In particular, two of the three mass differences between the isospin multiplets $N$, $\Lambda$, $\Sigma$, $\Xi$ determine the third --  the familiar Gell-Mann-Okubo formula, which works remarkably well. Also, all of the matrix elements of the perturbation can be expressed in terms of the baryon masses. In particular, to first order in symmetry breaking, the matrix element
\be\label{eq:sigma0} \sigma_0\equiv \frac{\hat{m}}{2M_N}\langle N(p)| \;\ubar u + \dbar d -2\,\sbar s\,|N(p)\rangle\ee
is determined by the masses of the baryon octet:
\be\label{eq:sigma0 SU(3)}  \sigma_0=(M_\Xi+M_\Sigma-2M_N)/(S-1)\{1+O(m_s-\hat{m})\}\,,\ee
with $S\equiv m_s/\hat{m}$. The value of $S$ is well determined by the work done on the lattice: $S=27.46(44)$ \cite{FLAG2014}. With the observed baryon mass values, the formula \eqref{eq:sigma0 SU(3)} gives $\sigma_0\simeq 25$ MeV (the precise number depends on how the isospin breaking effects are accounted for and whether one uses linear mass formulae or quadratic ones -- at first order in symmetry breaking, these apply equally well). The comparison with the result of the Roy-Steiner analysis, $\sigma_N=59.1(3.5)$ MeV, shows that -- if SU(3) does represent a decent approximate symmetry for the matrix elements of the operator $\qbar\lambda^{\hspace{-0.2em}a} q$, so that the leading order formula \eqref{eq:sigma0 SU(3)} for $\sigma_0$ only receives modest corrections -- the contribution from the strange quarks must reduce the one from $\ubar u+\dbar d$ by about a factor of two, in flat contradiction with the OZI-rule, which implies that the nucleon expectation value of $\sbar s$ is small. 

The contradiction involves three independent sources of information: $\pi N$ scattering, pionic atoms and masses of the baryon octet. While the estimate obtained for the correction $\Delta_R$ in the low energy theorem \eqref{eq:LET} relies on SU(2)$_{\indR}\times$SU(2)$_{\indL}$, the one for  $\sigma_0$ is based on SU(3). Since the symmetry breaking parameter $m_s-\hat{m}$ of SU(3) is about 26 times larger than the parameter $\hat{m}$ that measures the strength of SU(2)$_{\indR}\times$SU(2)$_{\indL}$ breaking, the higher order contributions to $\sigma_0$ are expected to be more important than the one from $\Delta_R$. They were studied in \cite{Gasser 1981} and were found to be substantial, on account of the contributions from the one-loop graphs of $\chi$PT, which are not analytic in the quark masses and strongly break SU(3) symmetry. Qualitatively, this can be understood, because the masses of the Nambu-Goldstone bosons which run around in these graphs are very strongly affected by symmetry breaking: the kaons are much heavier than the pions. The higher order contributions did not indicate a breakdown of SU(3), however: they were estimated to increase the matrix element defined in \eqref{eq:sigma0} from $\sigma_0=25$ MeV to $\sigma_0=35(5)$ MeV \cite{Gasser 1981}. The value $\sigma_N\simeq 45$ MeV obtained from the Karlsruhe-Helsinki partial waves then required a violation of the OZI-rule that looked acceptable: $y\equiv \langle N(p)|\,2\;\sbar s|N(p)\rangle/\langle N(p)|\, \ubar u + \dbar d |N(p)\rangle\simeq 0.2$ \cite{Gasser:1990ce}.  

The work done on the lattice indicates that the nucleon matrix element of $\sbar s$ is indeed small, confirming the second one of the two assumptions formulated at the beginning of this section. If that is so, the value found for $\sigma_N$ on the basis of (a) the Roy-Steiner equations, (b) the scattering lengths extracted from the pionic atom results and (c) the partial wave analysis of SAID then strongly violates the first one:  since the 'corrections' of order $(m_s-\hat{m})^2$ or higher must then more than double those from the 'leading' term of order $m_s-\hat{m}$, the part of the QCD Lagrangian that breaks SU(3), $\frac{1}{3}(m_s-\hat{m})(\ubar u + \dbar d-2\,\sbar s)$, can then not be treated as a perturbation.  In view of this, it is a mystery that the Gell-Mann-Okubo formula works so well. This formula also neglects contributions beyond first order and works in other multiplets as well -- it provided the basis for Gell-Mann's prediction of the mass of the $\Omega^-$ \ldots\,
It is difficult to understand how the approximate SU(3)-symmetry which explains the observed pattern of the hadron masses can miserably fail for the matrix elements of the operator relevant for the breaking of this symmetry. I add a few remarks directed towards a resolution of the puzzle.

\underline{Lattice}. Concerning the value of $\sigma_N$, the lattice approach can provide an excellent check: the $\sigma$-term concerns the manner in which the proton mass changes when the quark masses are varied. The lattice approach can also provide an accurate determination  of the nucleon expectation value of $\sbar s$ and thereby determine the matrix element $\sigma_0$. As briefly discussed in the review of Claude Bernard \cite{Bernard talk}, the available lattice data concerning the dependence of the nucleon mass on the masses of the quarks are difficult to understand as they superficially indicate a growth in proportion to the first power of $M_\pi$, which is not consistent with the theoretical understanding of the low energy structure. I do not doubt that the steady progress achieved in the lattice approach will eventually provide us with reliable and accurate values, not only for $\sigma_N$ and $\sigma_0$, but also for the nucleon matrix elements of the operator $\frac{1}{2}(m_u-m_d)(\ubar u- \dbar d)$,  which is responsible for the breaking of isospin symmetry in QCD and belongs to the same representation of SU(3) as the term $\frac{1}{3}(ms-\hat{m})(\ubar u+\dbar d-2\,\sbar s)$ which generates the breaking of SU(3)-symmetry.\footnote{After the closing of the workshop, new lattice results of the BMW-Collaboration for the nucleon matrix elements of $\ubar u$, $\dbar d$ and $\sbar s$ became available  \cite{BMW 2015}. Within errors, they confirm the picture drawn in \cite{Gasser 1981,Gasser:1990ce}. In particular, the va\-lues for the relative size of the proton matrix elements, $ \langle N(p)|\,\ubar u |N(p) \rangle/ \langle  N(p)|\, \dbar d |N(p)\rangle=1.20(3)(3)$, $y =0.20(8)(8)$, agree remarkably well with the old numbers  (the first and second errors indicate the statistical and systematic uncertainties, respectively). For the $\sigma$-term, the BMW result reads  $\sigma_N=38(3) (3) \,\mbox{MeV}$ and the value of $y$ then implies  $\sigma_0=30(4)(4)\,\mbox{MeV}$. For the part of the mass difference between neutron and proton that is due to the strong interaction, the BMW collaboration finds $M_{\mbox{\tiny QCD}}^{n-p}\equiv(m_d-m_u)\langle N(p)|\,\ubar u- \dbar d|N(p)\rangle/2M_N=2.52(17)(24)\,\mbox{MeV}$ \cite{Borsanyi:2014jba}, to be compared with the leading order SU(3) formula, $M_{\mbox{\tiny QCD}}^{n-p}=(M_\Xi-M_\Sigma)(m_d-m_u)/(m_s-\hat{m})\{1+O(m_s-\hat{m})\}$, which is analogous to \eqref{eq:sigma0 SU(3)} and gives 
$M_{\mbox{\tiny QCD}}^{n-p}\simeq 3.5\,\mbox{MeV}$.  For the coherence of the picture, it is essential that the corrections of $O(m_s-\hat{m})$ reduce the proton matrix element of $\ubar u-\dbar d$ but enhance the one of $\ubar u+ \dbar d-2\, \sbar s$, despite the fact that the two operators belong to the same irreducible representation of SU(3). The lattice results fully confirm the analysis of \cite{Gasser 1981} also in this regard. Since the effects of $O(m_s-\hat{m})$ are relatively large, the accuracy of the $\chi$PT calculation, which treats them as corrections, is limited -- the lattice approach is not subject to this limitation and will eventually arrive at a very sharp and detailed picture.

The new lattice results accentuate the puzzle discussed in my talk: the value of the $\sigma$-term in \eqref{eq:sigmaRS} differs from the result quoted in \cite{BMW 2015} by more than three standard deviations. The most interesting conclusion to draw would be that the data on $\pi N$ scattering are in conflict with QCD, but this looks somewhat premature \ldots} 

\underline{Data}. The $\sigma$-term represents a small contribution to the scattering amplitude and it is not easy to reliably fish it out from the measured cross sections. There are notorious discrepancies in the data on $\pi N$ scattering. Some of the charge exchange data, for instance, are difficult to reconcile with those on the elastic channels. H\"ohler and coworkers had tested their representation with partial wave dispersion relations and partial wave relations and found satisfactory consistency  \cite{Sainio:2009zz}. The SAID analysis includes many more data but was not subject to these tests \cite{Mikko}. A direct comparison of the Roy-Steiner analysis with the available experimental information about  $\pi N$ scattering and pionic atoms is called for. Is it possible to reliably estimate the uncertainties to be attached to the Roy-Steiner representation of the scattering amplitude ? 

\underline{$\chi$PT}.  According to \cite{Alarcon:2011zs,Ren:2014vea}, $\chi$PT can accommodate a large $\sigma$-term together with a small violation of the OZI-rule (see the talk by Xiu-Lei Ren \cite{Ren talk}). 
It would be very instructive to sort out the pattern of SU(3)-breaking on this basis. What is the mass of the octet if the three light quarks are given the same mass, $\bar{m}=\frac{1}{3}(m_u+m_d+m_s)$ ? What fraction of the splitting between the isospin multiplets  is due to the perturbation of $O(m_s-\hat{m})$, what is the remainder due to the higher order contributions ? How large are the SU(3)-violations in the matrix elements of the operators $\ubar u+\dbar d+\sbar s$,\, $\ubar u+\dbar d-2\,\sbar s$ and $\ubar u-\dbar d$\,?  Does this explain why first order perturbation theory works in one case but fails in the other ?

\underline{Isospin breaking}. As mentioned above, the symmetry properties of the vacuum protect the pions from isospin breaking. Weinberg pointed out that the nucleons are not protected -- in the nucleon matrix elements, the effects generated by the fact that $m_u$ is different from $m_d$  are inherently stronger \cite{Weinberg 1977}. It is important to explore these, not only theoretically but also experimentally   \cite{Bernstein:1998ip,Bernstein:2013aia}. 
A coherent dispersive analysis of isospin breaking in the $\pi N$ scattering amplitude must account for the fact that the $\pi N$ coupling constant, which parameterizes the dominating contributions at low energies, is not isospin invariant. For the pionic atoms, isospin breaking has carefully been analyzed \cite{Gasser:2007zt}. The isospin asymmetries in the $\pi N$ scattering lengths have been investigated within $\chi$PT \cite{Hoferichter:2009ez} and the determination of the $\sigma$-term based on the Roy-Steiner equations also accounts for isospin breaking \cite{Hoferichter:2015dsa,Hoferichter:2015hva}. A reliable determination of the nucleon matrix elements of the operator  $\frac{1}{2}(m_u-m_d)(\ubar u- \dbar d)$, which is responsible for the breaking of isospin symmetry in QCD, would 
be of considerable interest also for our understanding of the mass difference between proton and neutron. This is the theme I started with and now return to. 

\section{Cottingham formula}
As discussed in the introduction, the mass difference between proton and neutron does receive a contribution also from the electromagnetic interaction, albeit of a sign opposite to what is observed. To leading order, the electromagnetic self-energy of a proton or a neutron is given by an integral over a matrix element of the time-ordered product of two electromagnetic currents \cite{Cottingham}:
\be\label{eq:Mgamma1} M_\gamma=\frac{e^2}{4M_N}\!\int\hspace{-1mm} d^4\hspace{-0.5mm}x\; D_{\mu\nu}(x)\langle N(p)|T\,j^\mu(x)\,j^\nu(0)|N(p)\rangle\,,\ \ee
where $D_{\mu\nu}(x)$ is  the photon propagator.
The Fourier transform of $\langle N(p)|T\,j^\mu(x)\,j^\nu(0)|N(p)\rangle$ represents the amplitude for forward Compton scattering. In the integral in \eqref{eq:Mgamma1}, the photon is off-shell: the amplitude for {\it virtual} Compton scattering is relevant here.  
Lorentz invariance implies that the integral is independent of the spin direction of the particle, so that it suffices to know the spin-averaged scattering amplitude. As a consequence of current conservation, the spin-average only involves two invariants, which I denote by $T_1(\nu,q^2),T_2(\nu,q^2)$, where  $q$ is the four-momentum of the virtual photon exchanged in the process and $\nu=p\cdot q/M_N$ is the photon energy in the Lab frame. Expressed in terms of these, the Cottingham formula \eqref{eq:Mgamma1} takes the form
\be\label{eq:Mgamma2} M_\gamma  =\frac{-i\, e^2}{2M_N(2\pi)^4}\hspace{-1mm} \int\hspace{-1.5mm} \frac{d^4q}{q^2+i\epsilon} \{3 q^2 T_1(\nu,q^2)+(2\nu^2+q^2)T_2(\nu,q^2)\}\,.\ee 

The imaginary part of the scattering amplitude is determined by the Fourier transform of the current commutator matrix element $\langle N(p)|[j^\mu(x),j^\nu(0)]|N(p)\rangle$ and is related to the total cross section of electroproduction, $e+N\rightarrow e + \mbox{anything}$. Denoting the corresponding structure functions by $V_1(\nu,q^2), V_2(\nu,q^2)$, we have: 
\be\label{eq:DR}\mbox{Im}T_1(\nu,q^2)=\pi\, V_1(\nu,q^2)\,,\hspace{2em}\mbox{Im}T_2(\nu,q^2)=\pi\, V_2(\nu,q^2)\,.\ee 
 Since asymptotic freedom ensures that, at short distances, the quarks behave like free particles, the retarded amplitude $ \theta(x^0)\langle N(p)|[j^\mu(x),j^\nu(0)]|N(p)\rangle$ as well as the time-ordered one are unambiguously fixed by the matrix element of the current commutator. 

At short distances, the time-ordered product behaves like  $\langle N(p)|T\,j^\mu(x)\,j^\nu(0)|N(p)\rangle\propto 1/x^2$. Since the photon propagator also behaves like this, the integral in \eqref{eq:Mgamma1} diverges logarithmically, as in the case of the electron mass in QED. The divergence is absorbed in the e.m.~renormalization of the parameters $g,m_u,m_d,\ldots$ that occur in the QCD Lagrangian. Since only the operators  belonging to $m_u$ and $m_d$ carry isospin, only the renormalization of these parameters matters for the mass difference between proton and neutron. The renormalizations of $m_u$ and $m_d$ are proportional to $(\frac{2}{3}e)^2 m_u$ and $(-\frac{1}{3}e)^2m_d$, respectively. Accordingly, the coefficient of the logarithmic divergence in $M_\gamma^p-M_\gamma^n$ is proportional to the proton matrix element of the operator $e^2(4m_u-m_d)\,(\ubar u -\dbar d)$: in the chiral limit,  the e.m.~mass difference between proton and neutron is finite. In reality there is a logarithmic divergence, but the coefficient is tiny.

The asymptotic behaviour of the amplitudes for large values of $\nu$ and fixed $q^2$ has extensively been studied in perturbative QCD. It has been shown that this theory "reggeizes", so that the behaviour can be analyzed in the framework of Reggeon field theory \cite{Reggeons}. The exchange of a Reggeon with Regge trajectory $\alpha(t)$ generates contributions of the type $T_1(\nu,q^2)\sim \nu^\alpha$, $T_2(\nu,q^2)\sim \nu^{\alpha-2}$, where $\alpha=\alpha(0)$ is the intercept of the trajectory. There is solid experimental evidence for the presence of Reggeons also in the data. Since the intercepts all obey  $\alpha<2$ (presumably, the highest one, the Pomeron, corresponds to a branch point at $\alpha=1$), Regge behaviour ensures that $T_2(\nu,q^2)$ obeys an unsubtracted dispersion relation at fixed $q^2$, while $T_1(\nu,q^2)$ requires a subtraction. Since $V_1(\nu,q^2)$ and $V_2(\nu,q^2)$ are odd in $\nu$, the dispersion relations may be written in the form:  
\be \label{eq:disp}T_1(\nu,q^2)=S_1(q^2)+2\nu^2\hspace{-1.5mm}\int _{\!0}^\infty\hspace{-0.5mm}\frac{d\nu'}{\nu'} \,\frac{V_1(\nu',q^2)}{\nu'^2-\nu^2-i\epsilon}\,,\hspace{1em}
 T_2(\nu,q^2)=2\hspace{-1mm} \int _{\!0}^\infty\hspace{-1mm}d\nu'\nu'\frac{V_2(\nu',q^2)}{\nu'^2-\nu^2-i\epsilon}\,.\ee 

\section{Reggeons and fixed poles} 

The structure functions are directly measurable only in the space-like region, $q^2\leq 0$, but a beautiful theorem due to Jost, Lehmann \cite{JostLehmann} and Dyson \cite{Dyson} states that causality (the fact that the current commutator vanishes outside the light-cone) determines their continuation into the time-like region almost uniquely: the continuation is unique up to polynomials in the variable $\nu$. Hence the scattering amplitudes $T_1(\nu,q^2)$ and $T_2(\nu,q^2)$ are uniquely determined by the cross section of electroproduction, up to an ambiguity of the form
\be\label{eq:DeltaT} \Delta T_i(\nu,q^2)=\sum_{n=0}^N c_i^n(q^2)\nu^{2n}\hspace{1em}(i=1,2)\,,\ee
where the imaginary parts of the coefficients $c_i^n(q^2)$ vanish in the space-like region. In Regge pole language, integer powers of $\nu$ represent {\it fixed poles}: unlike regular Reggeons, whose position in the angular momentum plane depends on the momentum transfer between the particles involved in the collision, a term of this type does not move along a trajectory. Regge behaviour rules out fixed poles in $T_2$, but in $T_1$, a term with $n=0$  is not a priori excluded. 

In \cite{GL 1975}, it is assumed that the asymptotic behaviour of the virtual Compton scattering amplitude can be understood in terms of Reggeon exchange and that a fixed pole does not occur.  In the following, I refer to this assumption as the {\it Reggeon dominance hypothesis} \cite{GHLR}.  It implies that the functions $T_1(\nu,q^2)$, $T_2(\nu,q^2)$ are fully determined by the values of the structure functions $V_1(\nu,q^2),V_2(\nu,q^2)$ in the space-like region, i.e.~by the electroproduction cross section. In particular, the subtraction function $S_1(q^2)$ in the dispersion relation \eqref{eq:disp} for $T_1(\nu,q^2)$ does then not represent a quantity that is independent of the structure functions, but is determined by these. Likewise, the Cottingham formula then fixes the electromagnetic contribution to the mass difference between proton and neutron in terms of the cross section for electroproduction, thereby allowing the evaluation of this formula on the basis of experiment, despite the need of a subtraction. The numerical result obtained in \cite{GL 1975} with the experimental information about the cross sections available at the time is $M^p_\gamma-M^n_\gamma=0.76(30)$ MeV.  The data were consistent with the scaling laws of Bjorken, which were used to evaluate the contributions from the deep inelastic region. These turned out to be too small to stick out from the uncertainties of the calculation. 

The assumption that the matrix elements of the current commutator are free of fixed poles is by no means generally accepted, however. For a discussion of fixed poles in the framework of Regge Theory, see the textbook~\cite{Collins:1977jy}. Even before the advent of QCD,  the possible presence of such contributions was discussed in the literature (see, e.g.~\cite{Damashek:1969xj,Zee:1972nq,Creutz:1973zf,Schierholz:1973ix}). More recently, the universality conjecture formulated in~\cite{Brodsky:2008qu} has received considerable attention (see e.g.~\cite{Muller:2015vha} and the papers quoted therein). To my knowledge, the question of whether or not the Reggeon contributions fully account for the high energy behaviour of the Compton amplitude at fixed photon virtuality remains open. If the answer should turn out to be negative, that would be most interesting, as it would imply that our understanding of the asymptotic behaviour of QCD is inherently incomplete: what is the origin of the additional contributions and how can they be determined experimentally ? I only add two comments concerning this issue.

1. The concept of 'fixed pole' does not always refer to polynomial contributions of the form \eqref{eq:DeltaT}. The 'fixed pole' term investigated in the work of Damashek and Gilman \cite{Damashek:1969xj}, for instance, does not concern $T_1$ at all, but represents a contribution to $T_2$, which asympotically falls off in proportion to $1/\nu^2$. They consider real Compton scattering, $q^2=0$, and work with the amplitude $f_1(\nu)=\alpha_{em}\nu^2T_2(\nu,0)/M_N$. As they point out, the asymptotic behaviour of  $f_1(\nu)$ is not fully accounted for by the contributions from Reggeon exchange: denoting the latter by $f_1^R(\nu)$, the difference between $f_1(\nu)$ and $f_1^R(\nu)$ does not tend to zero when $\nu$ tends to infinity, but approaches a constant,  $f_1(\nu)-f_1^R(\nu)\rightarrow C$. The authors refer to $C$ as a fixed pole contribution. 

The fixed pole contributions permitted by the Jost-Lehmann-Dyson theorem are of different nature. As mentioned above, Regge behaviour implies that $T_2(\nu,0)$ cannot contain a fixed pole of the type \eqref{eq:DeltaT}. This does not prevent the amplitude $\nu^2 T_2(\nu,0)$ from containing a constant term in the asymptotic behaviour. Quite to the contrary, even if $V_2$ would tend to zero so rapidly that the integral $\int_0^\infty d\nu' \nu' V_2$ converges, the representation \eqref{eq:DR} would imply that $\nu^2T_2(\nu,0)$ then tends to a constant: causality does require the occurrence of 'fixed poles' of the type considered by Damashek and Gilman, but this is not in conflict with the Reggeon dominance hypothesis, nor does it touch the issue of whether or not the electroproduction cross section unambiguously determines the difference between the electromagnetic self-energies of proton and neutron. 

2. The short distance properties of QCD ensure that, if both $\nu$ and $q^2$ are large, the behaviour of $T_1(\nu,q^2)$ and $T_2(\nu,q^2)$ is governed by the perturbative expansion in powers of the strong coupling constant, so that it is meaningul to investigate the contributions from individual graphs. The behaviour in the Regge region, where only $\nu$ becomes large while the virtuality $q^2$ is kept fixed, is a much more complex affair that is not governed by the short distance properties of QCD. In particular, values of $q^2$ of the order of $\Lambda_{\,\mbox{\tiny QCD}}^2$ are outside the reach of perturbation theory, even if $\nu$ is large. An infinite set of graphs needs to be summed up to understand the high-energy behaviour of the amplitudes in the Regge region. Possibly QCD reggeizes only partially -- if it should turn out that, in $T_1$, the remainder does contain a fixed pole, then it ought to be possible to identify this contribution explicitly, so that it can be accounted for, in particular also in the Cottingham formula.  

\section{Recent work on the Cottingham formula}

Recently, the Cottingham formula was reexamined \cite{WCM}.\footnote{Note that the claims made in that reference about the analysis in \cite{GL 1975} are wrong; they are rectified in \cite{GHLR}.} The authors observe that the value of the subtraction function at $q^2=0$ is related to the magnetic polarizability of proton and neutron (what counts in connection with the Cottingham formula is the difference between the subtraction functions relevant for proton and neutron). They estimate the value of $S_1^{p-n}(0)$ with the experimental information about the polarizabilities. As information about the dependence of the subtraction function on $q^2$ is not at their disposal, the authors make a simple ansatz for that and come up with $M^p_\gamma-M^n_\gamma=1.30(3)(47)$ MeV, substantially higher than the old estimate quoted above.  As pointed out in \cite{Thomas:2014dxa}, the ansatz used in \cite{WCM} is not consistent with the short distance properties of QCD -- the corresponding coefficient of the logarithmic divergence is too large. The deficiency is repaired in \cite{Erben:2014hza} and the evaluation of the remaining contributions is confirmed within errors. The net result obtained with the improved ansatz  is $M^p_\gamma-M^n_\gamma=1.04(35)$ MeV. 

It is instructive to compare the calculations of \cite{WCM}, \cite{Erben:2014hza} with the analysis of \cite{GL 1975}, which is based on Reggeon dominance. This is done in \cite{GHLR}, with the following result:

\vspace{0.4em}
\noindent\hspace{1em}\parbox{14.8cm}{\hspace{-0.8em}$\bullet\,$ If the ansatz made in \cite{WCM} is replaced by the subtraction function that follows from Reggeon dominance, while all other elements of the calculation are left as they are, the central value drops to $M^p_\gamma-M^n_\gamma=0.63$ MeV.

\hspace{-0.8em}$\bullet\,$ Repeating the exercise with the alternative ansatz made in \cite{Erben:2014hza} leads to $M^p_\gamma-M^n_\gamma=0.67$ MeV. }

\vspace{0.2em}
\noindent In either case, the old estimate, $M^p_\gamma-M^n_\gamma=0.76(30)$ MeV \cite{GL 1975} is thus confirmed: as far as those contributions to the Cottingham formula that do not come from the subtraction function are concerned, the data acquired in the course of the last 40 years reduce the uncertainties but do not indicate that the central value must be revised significantly. The reason why the numbers  obtained in \cite{WCM,Erben:2014hza} deviate from the one given in \cite{GL 1975} is that the authors replace the subtraction function obtained from Reggeon dominance with an ansatz of their own. 
The main problem with these calculations is the systematic theoretical error -- I do not know of a method that would allow one to estimate the uncertainty to be attached to an ansatz. 

The renormalized version of the Cottingham formula derived in \cite{GL 1975} relies on Bjorken scaling and does not account for the scaling violations, which in the meantime have thoroughly been explored, both theoretically and experimentally. Today, representations of the structure functions are available that are consistent, not only with the data on electroproduction, but also with the constraints imposed by perturbation theory, for the proton as well as for the neutron \cite{MRST,ABM13,ABM14}. These constraints imply that the two contributions occurring in the subtracted Cottingham formula (subtracted dispersion integral over $V_1$ and unsubtracted integral over $V_2$) both diverge, but the sum over all of the contributions, including the one from the subtraction function, is unambiguous and finite -- the divergences are absorbed in the e.m.~renormalization of $m_u$ and $m_d$. Unfortunately, in \cite{WCM,Erben:2014hza}, the contributions from the region where the photon virtuality $Q^2=-q^2$ becomes large are discarded: the integrals are cut off at $Q^2=2.0\pm 0.5\,\mbox{GeV}^2$. Since the dominating contributions from the deep inelastic region are absorbed in the e.m.~renormalization of $m_u$ and $m_d$, their net effect is expected to be small, but an evaluation of the subtracted Cottingham formula in the framework of QCD  is still missing. 

\section{Polarizabilities}

As mentioned above, the value of the subtraction function at $q^2=0$ is related to the polarizabilities of the nucleon. The status of our knowledge of these quantities is discussed in several talks given at this workshop \cite{Downie talk,Griesshammer talk,Feldman talk,Alarcon talk,Demissie talk,Beane talk}. Since Reggeon dominance determines the subtraction function, it also leads to a prediction for the difference between the polarizabilities of proton and neutron. The result for the difference between the electric polarizabilities reads\footnote{The numerical values given for the polarizabilities refer to the standard units, $10^{-4}\,\mbox{fm}^3$} $\alpha_E^p-\alpha_E^n=-1.7(4)$ \cite{GHLR}, 
consistent with the current experimental value $-0.9(1.6)$ \cite{Griesshammer} and somewhat more precise. The results obtained from the Baldin sum rule \cite{Baldin} then determine the difference of the magnetic polarizabilities: $\beta_M^p-\beta_M^n=0.3(7)$. Using the comparatively rather precise experimental results for the polarizabilities of the proton, an estimate for the polarizabilities of the neutron also follows: $\alpha_E^n=12.3(7)$, 
$\beta_M^n=2.9(9)$, numbers that are perfectly consistent with the experimental values $\alpha_E^n = 11.55(1.50)$, $\beta_M^n=3.65(1.50)$ \cite{Griesshammer}.  

 The fact that the results obtained from Reggeon dominance are consistent with experiment  amounts to a nontrivial test of the hypothesis that the Compton amplitude is free of fixed poles. Quite apart from the possibility of taking new data at small photon virtuality, an improved representation of the available experimental information on the cross sections in the intermediate energy region ($1.5\,\mbox{GeV}<W<3\,\mbox{GeV}$) is called for -- this would reduce the uncertainties in the prediction quite substantially (the shortcomings of the parameterizations available in that region were pointed out in \cite{Erben:2014hza}; for a detailed discussion, see \cite{GHLR}). Needless to say that a more accurate determination of the neutron polarizabilities would be most welcome, as it would sharpen the experimental test of the prediction. 
 
Note that the polarizabilites do not determine the subtraction function and do therefore not play any role in theoretical  determinations of the proton-neutron mass difference. In the papers discussed above, a result for the mass difference is obtained by bridging lack of knowledge with an ansatz, but it is clear that the question of whether or not the current commutator contains a fixed pole cannot be answered by making an ansatz. 

The main problem faced in the numerical evaluation of the subtraction function relevant for the difference between proton and neutron is that all of the well-established features of electroproduction drop out when taking the difference between proton and neutron: the leading terms of the chiral perturbation series are the same, the contribution from the most prominent resonance, the $\Delta(1232)$, is the same, and the leading asymptotic term due to Pomeron exchange is also the same. Since all of these contributions cancel out, not much is left over -- even the logarithmic divergences nearly cancel. Only a fixed pole could prevent the subtraction function relevant for the difference between proton and neutron from being small. The available data do not exclude the phenomenon, but indicate that, if a fixed pole does occur, then its residue must be small. 

\section{Summary and conclusion}
Several different methods are used to investigate the low energy structure of the Standard Model: experiment, $\chi$PT, dispersion theory, lattice approach, QCD sum rules, \ldots\, The theoretical analysis heavily relies on the symmetry properties of QCD. In this context, the light quark masses play an important role: chiral symmetry strictly holds only if they are set equal to zero. In the real world, the symmetry is broken -- the quark masses measure the strength of the symmetry breaking. Since the lowest states in the meson sector represent Nambu-Goldstone bosons, the underlying hidden symmetry imposes strong constraints on their properties.  
Dispersion theory provides good control over the dependence of the various quantities of physical interest (form factors, scattering amplitudes) on the external momenta, but it does not shed any light on the sensitivity of these quantities to the masses of the light quarks. 

In the meson sector, the interplay between the different methods has led to a coherent framework, which leads to firm and accurate predictions concerning various quantities relevant in flavour physics, in particular also concerning physics beyond the Standard Model. The interaction among pions of low energy is very well understood. In particular, the properties of the scattering amplitude in the region of the lowest resonance, which carries the quantum numbers of the vacuum, are known to remarkable accuracy. The low energy theorems of chiral symmetry have passed stringent tests. 

In the baryon sector, on the other hand, a satisfactory understanding of the low energy structure is not yet achieved. There is very significant progress in dispersion theory, but contact with lattice work yet needs to be established. In my talk, I focused on two specific issues in this domain: the $\sigma$-term and the proton-neutron mass difference. 
A thorough update of the dispersive analysis of the Karlsruhe-Helsinki collaboration is now available, based on the Roy-Steiner equations. It indicates that the $\sigma$-term can be determined rather accurately from the available data on $\pi N$ scattering and pionic atoms. The result, however, is puzzling: the same approximation that leads to the very successful Gell-Mann-Okubo formula for the masses of the baryon octet yields a prediction for the $\sigma$-term that is in conflict with the value obtained from the Roy-Steiner analysis -- this puzzle yet needs to be solved. A determination of the relevant matrix elements on the lattice would help to clarify the situation.

I also briefly reviewed recent work on the proton-neutron mass difference. The central issue in this context was identified long ago: the electromagnetic contribution to the mass difference can be calculated in terms of the cross section for electroproduction if and only if the nucleon matrix element of the current commutator is free of fixed poles \cite{GL 1975}. The e.m.~part of the mass difference consists of a sum of two terms: an integral over the structure functions  ('subtracted Cottingham formula') and an integral over the subtraction function that occurs in the dispersive representation of the Compton scattering amplitude $T_1$. While the subtracted integrals can be evaluated on the basis of what is known, it is still an open issue whether the asymptotic behaviour of the Compton scattering amplitude at fixed photon virtuality is fully accounted for by Reggeon exchange ('Reggeon dominance hypothesis') or whether the subtraction function contains an additional contribution from a fixed pole -- that would be most interesting, as it would imply that our understanding of the asymptotic behaviour of QCD is inherently incomplete.

The available lattice determinations of $M_\gamma^p-M_\gamma^n$ are consistent with the estimate obtained on the basis of the Reggeon dominance hypothesis. A reduction of the uncertainties in the lattice data could strengthen this test quite substantially. The evaluation of the contributions to the subtracted Cottingham formula arising from the deep inelastic region need to be updated as well, using a representation of the data that is consistent with the constraints imposed by perturbation theory, so that the scaling violations are accounted for. This is yet to be done.

The data on the nucleon polarizabilities also offer a test. The available experimental results are consistent with Reggeon dominance, but, in view of the rather large experimental uncertainties, not only in the polarizabilities but also in the cross sections for electroproduction at small photon virtuality, only a fixed pole with sizable residue is ruled out.  

\section*{Acknowledgments}
I thank Balasubramanian Ananthanarayan, Irinel Caprini, Gilberto Colangelo, J\"urg Gasser, Martin Hoferichter, Bastian Kubis, Jos\'e Pel\'aez, Akaki Rusetsky, Mikko Sainio and Peter Stoffer for useful information and comments. Also, I wish to thank Laura Marcucci and Michele Viviani for creating a most pleasant environment for the workshop.

\end{document}